\pdfoutput=1	

\documentclass{elsart}
\usepackage{graphicx}
\usepackage{amsmath}
\usepackage{amssymb}

\usepackage{amsopn} 
\DeclareMathOperator{\sgn}{sgn}

\begin{document}

\newcommand{\doi}[1] {doi: \href{http://dx.doi.org/#1}{#1}}

\newcommand{\rmi}{\mathrm{i}}
\newcommand{\rmd}{\mathrm{d}}
\newcommand{\bi}[1]{\mathbf{#1}}



\newcommand{\change}[2]{#1}
\newcommand{\addition}[1]{#1}

\newcommand{\detail}[1]{#1}


\begin{frontmatter}

\title{Quantifying metarefraction with confocal lenslet arrays}

\author{Tautvydas Maceina, Gediminas Juzeli\={u}nas}

\address{Institute of Theoretical Physics and Astronomy, Vilnius University, A. Go\"stauto 12, LT-01108 Vilnius, Lithuania}

\author{Johannes Courtial}

\address{School of Physics \& Astronomy, University of Glasgow, Glasgow G12~8QQ, United Kingdom}

\ead{johannes.courtial@glasgow.ac.uk}

\begin{abstract}
METATOYs can change the direction of light in ways that appear to, but do not actually, contravene the laws of wave optics.
This direction change applies only to part of the transmitted light beam; the remainder gets re-directed differently.
For a specific example, namely confocal pairs of rectangular lenslet arrays with no dead area between lenslets, we calculate here the fractions of power of a uniform-intensity light beam incident from a specific (but arbitrary) direction that get re-directed in different ways, and we derive an equation describing this redirection.
This will facilitate assessment of the suitability of METATOYs for applications such as solar concentration.
Finally, we discuss similarities  between the multiple refraction of light at the lenslet arrays and multiple refraction and reflection of cold atoms at a barrier in the presence of the light fields.
\end{abstract}



\end{frontmatter}

\section{Introduction}

METATOYs \cite{Hamilton-Courtial-2009} can be described as windows that ``refract'' (change the direction of) transmitted light rays.
This direction change can be, for example, a rotation around the window normal \cite{Hamilton-et-al-2009}, or indeed around any other direction \cite{Hamilton-et-al-2010};
a flipping (sign change) of one of the transverse direction components \cite{Hamilton-Courtial-2008a};
negative refraction that leads to pseudoscopic imaging \cite{Courtial-Nelson-2008};
and a variation of Snell's law in which sines are replaced by tangents \cite{Courtial-2008a} (and a generalisation thereof \cite{Hamilton-Courtial-2009b}).
It can be shown that most of these light-ray-direction changes can lead to light-ray fields that cannot be represented wave-optically \cite{Hamilton-Courtial-2009,Courtial-et-al-2011}.
However, this is not actually the case, and this apparent conflict with wave optics is resolved by introducing discontinuities into the wave front \cite{Tyc-et-al-2011}.

The wave-front discontinuities introduced by METATOYs  result in only part of the light being redirected as advertised; the remainder undergoes a different direction change.
This imperfection can be remedied by more careful optical design, for example insertion of arrays of field lenses into the common focal plane of confocal lenslet arrays (CLAs) \cite{Stevens-Harvey-2002}, which are examples of METATOYs \cite{Courtial-2008a}.
Nevertheless, it compromises the performance of METATOYs as components in visual optical instruments and in instruments for light-shaping applications.

One timely application of light shaping is solar concentration \cite{Coffey-2011}.
Testing whether or not METATOYs can offer anything new in this important area is an obvious task, one for which it is crucial to understand how much light undergoes what direction change.

\begin{figure}
\begin{center} \includegraphics{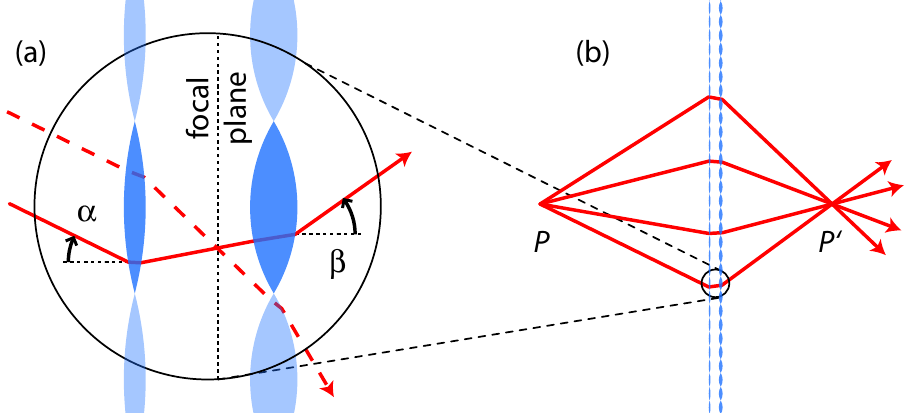} \end{center}
\caption{\label{CLAs-figure}Confocal lenslet arrays (CLAs).
The focal lengths of the lenslets in the left and right array are respectively $f_1$ and $f_2$.
(a)~Example of a light ray passing through corresponding lenslets (solid arrow) and non-corresponding lenslets (dashed arrow).
$\alpha$ and $\beta$ are the angles of incidence and refraction, respectively, of the former light ray.
(b)~Imaging properties of light rays that pass through corresponding lenslets in planar CLAs.}
\end{figure}

\begin{figure}
\begin{center} \includegraphics{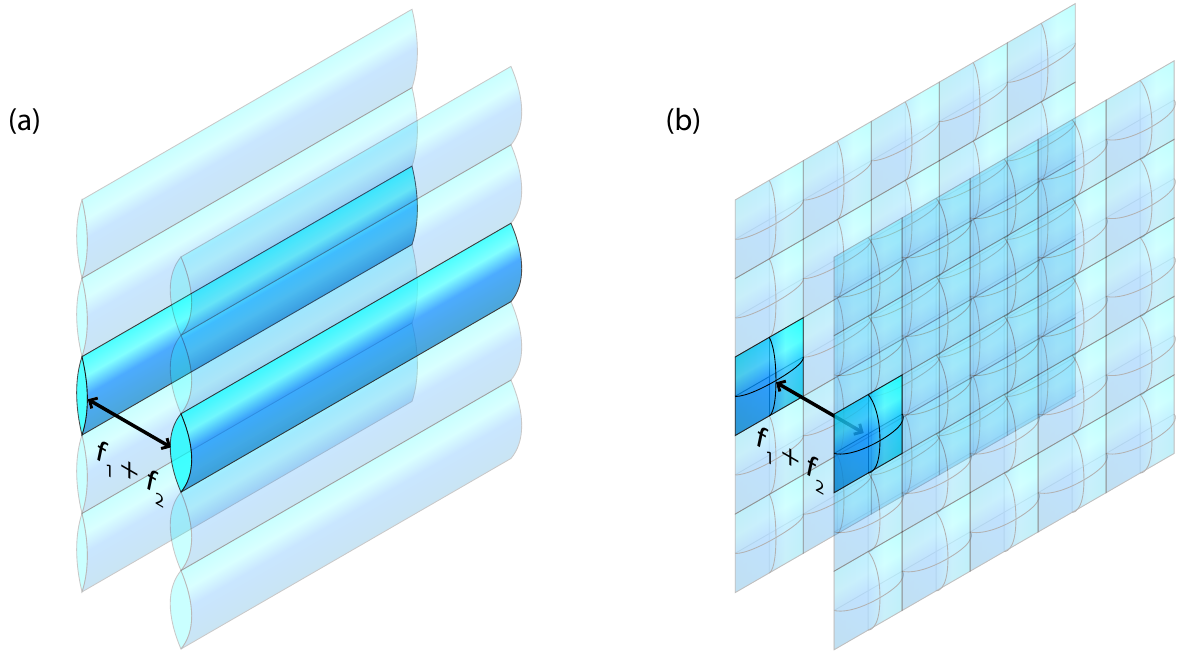} \end{center}
\caption{\label{CLAs-geometry-figure}Geometries of CLAs to which our calculation applies.
(a)~Confocal arrays of cylindrical lenslets;
(b)~confocal rectangular arrays of spherical lenslets with rectangular apertures.
In both cases, $f_1$ and $f_2$ is the focal length of the lenslets in the back array and the front array, respectively, and there is no dead area between neighbouring lenslets.
For clarity, a pair of lenslets that occupy corresponding positions in the two lenslet arrays are highlighted in both diagrams.}
\end{figure}

Here we address this point for simple CLAs \cite{Courtial-2008a} (Fig.\ \ref{CLAs-figure}).
Note that lenslet (or microlens) arrays are already being investigated as a way of improving the efficiency of photovoltaic cells (e.g.\ \cite{Tvingstedt-et-al-2008}).
We build on previous work on the mechanism behind, and qualitative effect of, different parts of a light beam being refracted differently.
Specifically, we calculate the fraction of a uniform-intensity light beam, incident from a specific (but arbitrary) direction, that undergoes the ``correct'' direction change upon transmission through the CLAs; we call this fraction $\zeta$.
We also calculate the fractions $\zeta_m$ of transmitted light that undergo other direction changes.
Our calculation is two-dimensional and directly describes confocal arrays of cylindrical lenslets that are invariant to translation in one direction (for example the $y$ direction; Fig.\ \ref{CLAs-geometry-figure}(a)), but the calculation also describes --- separately --- the relevant lateral projections\footnote{If the lenslet arrays are parallel to the $xy$ plane and periodic in the $x$ and $y$ directions, then the relevant lateral projections are into the $xz$ and $yz$ planes.} of light passing through rectangular arrays of lenslets with rectangular apertures (Fig.\ \ref{CLAs-geometry-figure}(b)).
If $\zeta_{xz}$ is the fraction of the power of the incident light that undergoes the correct direction change in the $xz$ projection, and $\zeta_{yz}$ is the corresponding fraction in the $yz$ projection, then the overall fraction of the power of incident light that undergoes the correct direction change is $\zeta_{xz} \zeta_{yz}$.
Our calculation makes approximations by assuming that there is no dead area between neighbouring lenslets, and that each lenslet redirects light rays like an ideal thin lens\footnote{This assumption implies that the individual lenslets have flat fields.}.
Our results are important as they can be used to answer the question whether or not CLAs, along with realising previously forbidden refraction qualitatively, can also overcome previously established \emph{quantitative} limits.

\section{Confocal lenslet arrays (CLAs)}

CLAs (Fig.\ \ref{CLAs-figure}) are formed by two parallel arrays of lenslets (or microlenses), separated by the sum of their focal lengths \cite{Courtial-2008a}.
In the simplest case, which we consider here, the lenslets' optical axes are perpendicular to both array planes, and each lenslet in one array has a corresponding lenslet in the other array with which it shares an optical axis.
Corresponding lenslets can be seen as tiny two-lens telescopes, so CLAs are simply arrays of telescopes. 
Note that there is a generalisation of CLAs \cite{Hamilton-Courtial-2009b}, but we do not consider it here.

When a light ray is transmitted through a telescope, its transverse positions on entering and on exiting the telescope are generally different.
This transverse offset occurs also in CLAs, but by miniaturising the telescopes the offset can be made small\footnote{The limit of useful miniaturisation has been exceeded when wave-optical effects begin to dominate.}.
Each telescope then acts like a pixel of the window formed by the CLAs; under the right conditions, the pixellation can be as unnoticeable as a computer monitor's.
That this approach works has been demonstrated experimentally~\cite{Courtial-et-al-2010}.

The change in direction of light rays that pass through a telescope consisting of lenses with focal lengths $f_1$ and $f_2$ can be described by the following equation, which describes the relationship between the angles of the light ray with the optical axis on the two sides of the telescope, $\alpha$ and $\beta$ (Fig.\ \ref{CLAs-figure}):
\begin{equation}
f_1 \tan \alpha = - f_2 \tan \beta.
\label{tan-refraction-equation}
\end{equation}
As transmission through CLAs is transmission through telescopes, CLAs therefore refract light rays according to this law of refraction, which is remarkably similar to Snell's law~\cite{Courtial-2008a}.

The law of refraction given by Eqn (\ref{tan-refraction-equation}) is interesting as it leads to perfect imaging (Fig.\ \ref{CLAs-figure}(b)) \cite{Courtial-2008a}.
Obviously, if it is realised experimentally with CLAs, then the imaging can only be as good as the offset is small, and as good as the lenslets redirect light like ideal thin lenses.
The case $f_1 = f_2$ corresponds to refraction between media with refractive indices of equal magnitude and opposite sign \cite{Veselago-1968,Pendry-2000}.

Strictly speaking, Eqn (\ref{tan-refraction-equation}) applies only to light rays that pass through corresponding lenslets, that is, light rays that exit the same telescope that they entered \cite{Courtial-2009}.
Provided this is the case, the direction change is independent of the precise position where the ray hits the first lenslet.
The direction change such light rays undergo is called standard refraction;
light rays that enter one lenslet and exit a non-corresponding lenslet (like the dashed light ray in Fig.\ \ref{CLAs-figure}(a)) undergo a different direction change called non-standard refraction~\cite{Courtial-2009}.

In this paper we consider only CLAs with a particularly simple geometry: the aperture width of each lenslet is the same in both arrays, and there is no dead area between neighbouring lenslets (i.e.\ the centre-to-centre separation between neighbouring lenslets equald the aperture width).
We describe such CLAs in terms of the dimensionless focal-length ratio \cite{Courtial-2008a},
\begin{align}
\eta = - \frac{f_2}{f_1},
\label{eta-equation}
\end{align}
and the f-number of the left lenslet,
\begin{align}
N = \frac{f_1}{2 r},
\label{N1-equation}
\end{align}
where $r$ is half the lenslets' aperture width.

Our analysis considers light travelling from left to right, and assumes that the effect of transmission through the two lenslets can be described by first taking into account the effect of the lenslet with focal length $f_1$ and then that of the lenslet with focal length $f_2$.
In other words, it assumes that the lenslet with focal length $f_1$ is to the left of the lenslet with focal length $f_2$, or that the two are in (or imaged into) the same plane, which is only the case if $f_1 + f_2 \geq 0$.
In terms of $\eta$ and $N$, this becomes
\begin{align}
N (1 - \eta) \geq 0
\label{right-order-condition}
\end{align}
(as $r > 0$, by definition).
The values of $\eta$ and $N$ 
are therefore restricted to combinations 
that satisfy this condition.


\section{\label{standard-refraction-section}Light undergoing standard refraction}

First we calculate the fraction of incident power that undergoes standard refraction.
For simplicity, we consider here (and in the following sections) the incident light to consist of parallel rays, all of the same brightness; the angle of incidence is $\alpha$.
Wave-optically, this is  a uniform plane wave.

\subsection{\label{case-1-section}The case $|\eta| \leq 1$} 

First we treat the case $|\eta| \leq 1$, which corresponds to $f_1 \geq |f_2|$.
For this case, condition (\ref{right-order-condition}) is satisfied provided that $N \geq 0$.
It is convenient to treat the ranges $-1 \leq \eta \leq 0$ and $0 \leq \eta \leq 1$ separately.

\begin{figure}
\begin{center} \includegraphics{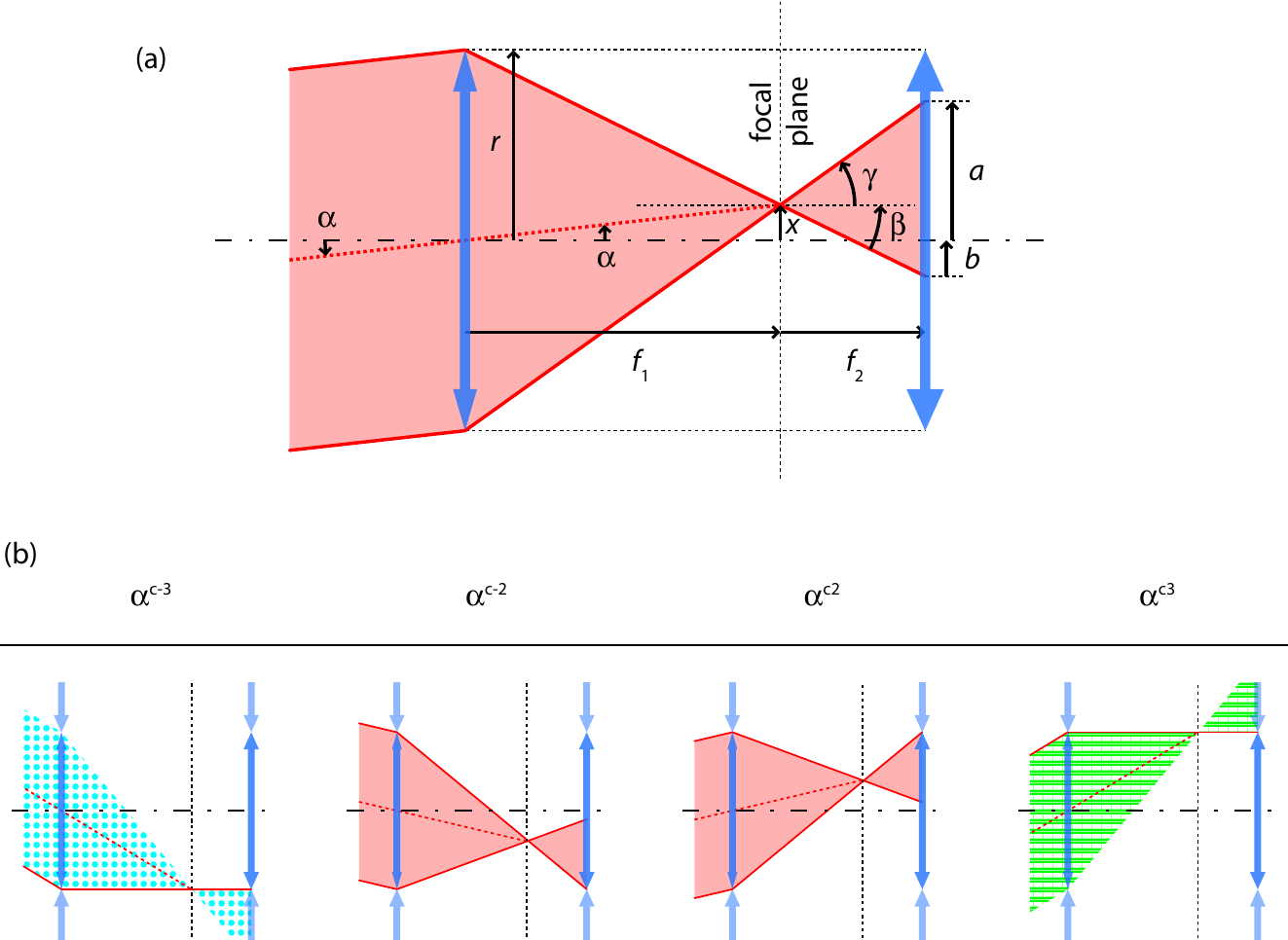} \end{center}
\caption{\label{case-1a-figure}(a)~Geometry of and (b)~critical beam paths through confocal lenslet arrays for the parameter range $-1 \leq \eta \leq 0$.
(a)~A collimated light beam enters a pair of corresponding lenslets from the left, inclined by an angle $\alpha$ with respect to the common optical axis.
The longitudinal cross-sectional area of the beam is shown in solid light red; its edges are marked by solid red lines.
The half-width of the lenslets is $r$.
\detail{Counter-clockwise angles are positive; upwards and rightwards distances are positive.}
(b)~Longitudinal cross-sectional
areas for beams incident at critical angles.
The 
part of the beam that passes through corresponding lenslets is filled in solid light red and its edges are marked by solid red lines.
The 
part of the beam that enters through the left lenslet and exits through the lenslet above (below) the corresponding right lenslet is filled with green horizontal stripes (turquoise polka dots).}
\end{figure}

\begin{figure}
\begin{center} \includegraphics{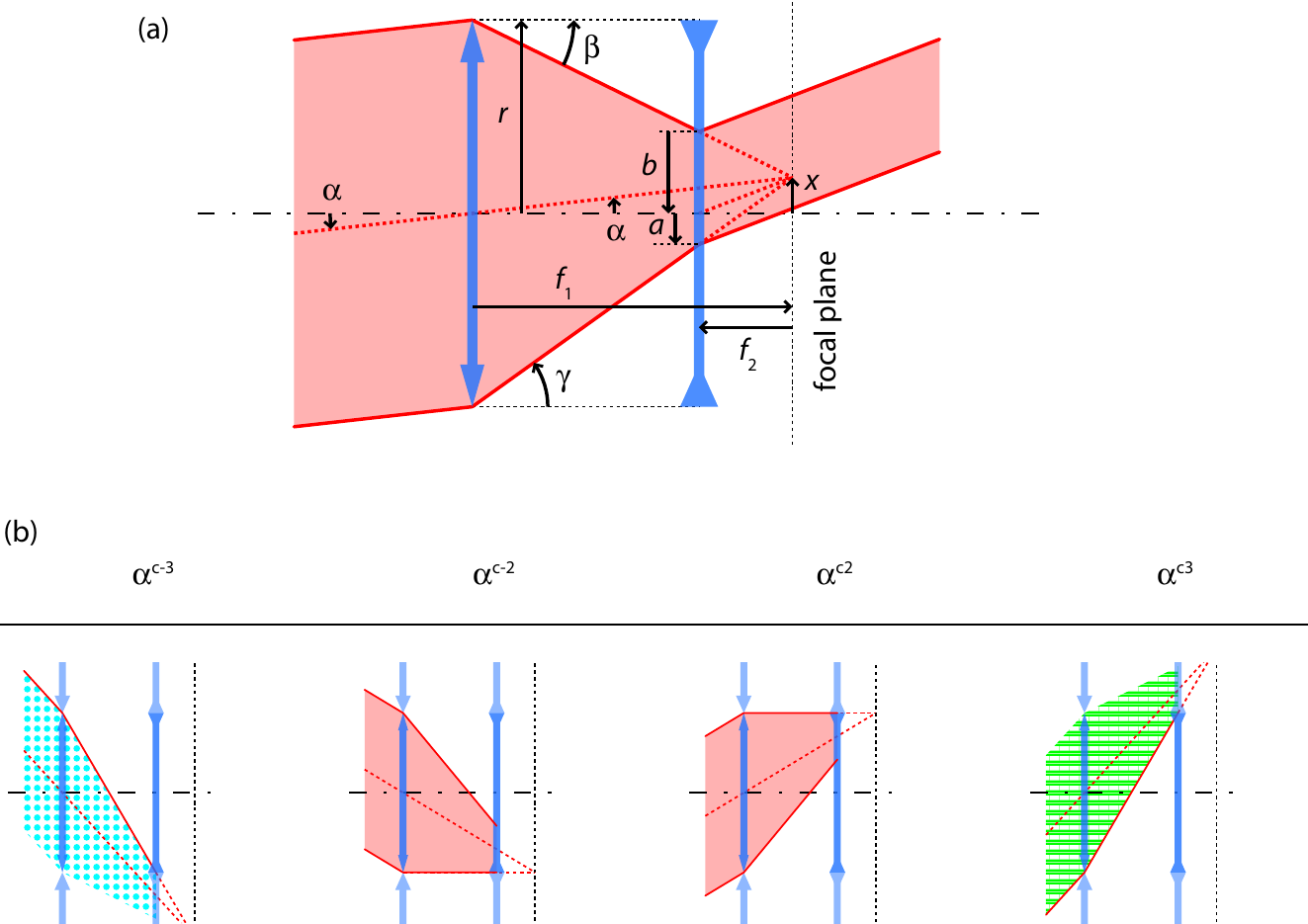} \end{center}
\caption{\label{case-1b-figure}(a)~Geometry of and (b)~critical beam paths through confocal lenslet arrays in the parameter range $0 \leq \eta \leq 1$.
See the caption of Fig.\ \ref{case-1a-figure} for further details.}
\end{figure}

Before calculating the fraction of the light that passes through the first (left) lenslet, which then passes through the corresponding (right) lenslet, we briefly discuss what we expect to happen for different values of the angle of incidence.
Key throughout is the dependence of the light in the plane of the right lenslet array, described by the parameters $a$ and $b$, on the angle of incidence, $\alpha$ (see Figs \ref{case-1a-figure}(a) and \ref{case-1b-figure}(a)).
To derive this relationship, we start with the equations
\begin{equation}
\tan \gamma = \frac{r + x}{f_1} = \frac{a - x}{f_2}, \qquad 
\tan \beta = \frac{r - x}{f_1} = \frac{b + x}{f_2}, 
\label{tan-gamma-tan-beta-eqn-1}
\end{equation}
and so
\begin{align}
a &= 
	\frac{r f_2 + x (f_1 + f_2)}{f_1}, \\
b &= 
	\frac{r f_2 - x (f_1 + f_2)}{f_1}.
\end{align}
With $x = f_1 \tan \alpha$, these expressions become the sought-for equations describing the dependence of $a$ and $b$ on $\alpha$:
\begin{align}
a &= r f_2 / f_1 + (f_1 + f_2) \tan \alpha, \label{a-alpha-case-1-eqn} \\
b &= r f_2 / f_1 - (f_1 + f_2) \tan \alpha. \label{b-alpha-case-1-eqn}
\end{align}

For normal incidence ($\alpha = 0^\circ$), $a$ and $b$ take on the value $a = b = r f_2 / f_1$.
Specifically, $|a|, |b| \leq r$, which means that all of the light that enters through the left lenslet exits again through the corresponding right lenslet.

As $\alpha$ is increased to a ``critical angle'' $\alpha^{c2}$, the upper edge of the beam hits the upper edge of the right lenslet.
In Ref.\ \cite{Courtial-2009}, $\alpha^{c2}$ is called the ``second critical angle of incidence''; we will encounter the other critical angles in due course.
For $-1 \leq \eta \leq 0$, $\alpha^{c2}$ is reached when $a = r$ (see Fig.\ \ref{case-1a-figure}(b)).
Substitution into Eqn (\ref{a-alpha-case-1-eqn}) gives
\begin{equation}
\tan \alpha^{c2} = \frac{r (f_1 - f_2)}{f_1 (f_1 + f_2)}
= \frac{1}{2 N} \frac{1+ \eta}{1 - \eta} \qquad (\eta \leq 0).
\label{alpha^{c2}-eta<0-equation}
\end{equation}
For $0 \leq \eta \leq 1$, $\alpha^{c2}$ is reached when $b = -r$ (see Fig.\ \ref{case-1b-figure}(b)); substitution into Eqn (\ref{b-alpha-case-1-eqn}) gives
\begin{equation}
\tan \alpha^{c2} = \frac{r}{f_1}
= \frac{1}{2 N} \qquad (\eta \geq 0).
\label{alpha^{c2}-eta>0-equation}
\end{equation}

Any further increase in $\alpha$ means that part of the beam misses the corresponding right lenslet.
Another critical angle, $\alpha^{c3}$ (the third critical angle of incidence, according to the definition in Ref.\ \cite{Courtial-2009}), is reached when the lower edge of the beam passes through the upper edge of the right lenslet.
For $-1 \leq \eta \leq 0$, this happens when $b = -r$ (see Fig.\ \ref{case-1a-figure}(b)); substitution into Eqn (\ref{b-alpha-case-1-eqn}) yields
\begin{equation}
\tan \alpha^{c3} = \frac{r (f_2 / f_1 + 1)}{f_1 + f_2} = \frac{r}{f_1} = \frac{1}{2 N} \qquad (\eta \leq 0).
\label{alpha^{c3}-eta<0-equation}
\end{equation}
For $0 \leq \eta \leq 1$, it happens when $a = r$ (see Fig.\ \ref{case-1b-figure}(b)); substitution into Eqn (\ref{a-alpha-case-1-eqn}) gives
\begin{equation}
\tan \alpha^{c3} = \frac{r (f_1 - f_2)}{f_1 (f_1 + f_2)} = \frac{1}{2 N} \frac{1+\eta}{1-\eta}
\qquad (\eta \geq 0).
\label{alpha^{c3}-eta>0-equation}
\end{equation}
For any value $\alpha \geq \alpha^{c3}$, none of the beam passes through the corresponding right lenslet.

It is perhaps worth noting that the expressions for $\alpha^{c3}$ for $-1 \leq \eta \leq 0$ (Eqn (\ref{alpha^{c3}-eta<0-equation})) and for $\alpha^{c2}$ for $0 \leq \eta \leq 1$ (Eqn (\ref{alpha^{c2}-eta>0-equation})) are identical, and so are the expressions for $\alpha^{c2}$ for $-1 \leq \eta \leq 0$ (Eqn (\ref{alpha^{c2}-eta<0-equation})) and for $\alpha^{c3}$ for $0 \leq \eta \leq 1$ (Eqn~(\ref{alpha^{c3}-eta>0-equation})).

So far we have only discussed positive angles of incidence.
What happens for negative values of $\alpha$?
For symmetry reasons, the same as what happens for positive values of $\alpha$ with the same modulus.
Starting again from normal incidence and this time decreasing the value of $\alpha$, the beam will again start to be clipped as another critical angle, $\alpha^{c-2}$, is reached.
If $\alpha$ is decreased further, the fraction of the beam intersecting the corresponding right lenslet decreases until it reaches zero, at the critical angle $\alpha^{c-3}$.
From the symmetry, and from Figs \ref{case-1a-figure}(b) and \ref{case-1b-figure}(b), it can be seen that
\begin{align}
\alpha^{c-2} = - \alpha^{c2}, \quad \alpha^{c-3} = - \alpha^{c3}.
\end{align}

Now we calculate the power fraction of a uniform plane-wave beam of light that passes through corresponding lenslets.
As before, the beam is incident from the left; we consider the part of the beam that has passed through a specific lenslet in the left lenslet array.
In the plane of the corresponding right lenslet, the height of the beam is $|a + b|$.
In the diagrams shown in Figs \ref{case-1a-figure}(a) and \ref{case-1b-figure}(a), this is also the part of the corresponding right lenslet that is illuminated by this beam.
The reason is that those diagrams are drawn for cases in which all light that passes through the left lenslet also passes through the right lenslet, that is, an angle of incidence $\alpha$ between $\alpha^{c-2}$ and $\alpha^{c2}$.
For angles of incidence outside this range, the height of the second lenslet that is illuminated by the beam is less than the height of the beam, as some of the beam now misses the second lenslet.
Generally, we can calculate the fraction $\zeta$ of power that undergoes standard refraction as the fraction of the height of the right lenslet that is illuminated by the beam and the height of the beam in the plane of the right lenslet, $|a+b|$.

For angles of incidence between $0^\circ$ and the second critical angle, $\alpha^{c2}$, the fraction $\zeta$ is simply 1;
above the third critical angle, $\alpha^{c3}$, $\zeta = 0$.
For angles of incidence between $\alpha^{c2}$ and $\alpha^{c3}$, the top of the beam in the plane of the right lenslet is above the top edge of that right lenslet.
The height of the beam illuminating the right lenslet is therefore $r+b$ in the case $-1 \leq \eta \leq 0$ and $r-a$ in the case $0 \leq \eta \leq 1$, and so
\begin{align}
\zeta = \frac{r+b}{a+b} 
&= \frac{ \eta - 1 - 2 N (\eta - 1) \tan \alpha }{2 \eta}
\qquad (-1 \leq \eta \leq 0), \\
\zeta = \frac{r-a}{-a-b}
&= \frac{ \eta + 1 + 2 N (\eta - 1) \tan \alpha }{2 \eta}
\qquad (0 \leq \eta \leq 1).
\end{align}
For negative angles of incidence, the fraction is the same as for a positive angle with the same modulus.
These results can be summarised as follows:
\begin{equation}
\zeta = \begin{cases}
1 & \mbox{if } |\alpha| \leq \alpha^{c2}, \\
\frac{ \eta + \sgn(\eta) [ 1 + 2 N (\eta - 1) \tan |\alpha| ] }{2 \eta}
& \mbox{if }
\alpha^{c2} \leq |\alpha| \leq \alpha^{c3}, \\
0 & \mbox{if } \alpha^{c3} \leq |\alpha|,
\end{cases}
\end{equation}
where $\sgn(x)$ is the signum function, i.e.\
\begin{align}
\sgn(x) = \left\{
\begin{array}{ll}
+1 & \mbox{if } x > 0, \\
0 & \mbox{if } x = 0, \\
-1 & \mbox{if } x < 0.
\end{array}
\right.
\end{align}

\begin{figure}
\begin{center} \includegraphics{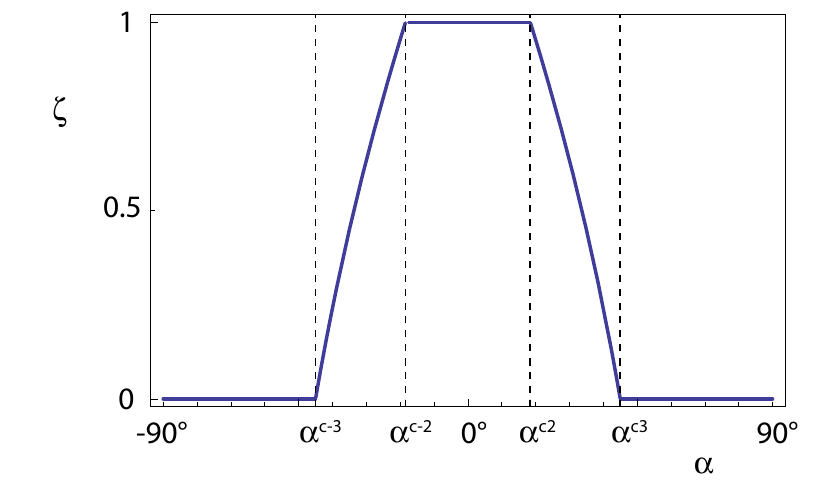} \end{center}
\caption{\label{zeta-alpha-graph-case1-example-figure}Example of a plot of $\zeta$ as a function of $\alpha$.
The graph is plotted for $\eta = -0.5$ and $N =0.5$ (e.g.\ $f_1 = 1$, $f_2 = 0.5$, $r = 1$).
The critical angles are indicated on the $\alpha$ axis; their values are $\alpha^{c2} = 18.4^\circ$ and $\alpha^{c3} = 45^\circ$.}
\end{figure}

Fig.\ \ref{zeta-alpha-graph-case1-example-figure} shows an example plot of $\zeta$ as a function of the angle of incidence, $\alpha$. 
The role of the critical angles $\alpha^{c2}$ and $\alpha^{c3}$ as can clearly be seen:  $\zeta = 1$ for $-\alpha^{c2} \leq \alpha \leq \alpha^{c2}$, and $\zeta = 0$ for $|\alpha| \geq \alpha^{c3}$.

\begin{figure}
\begin{center} \includegraphics{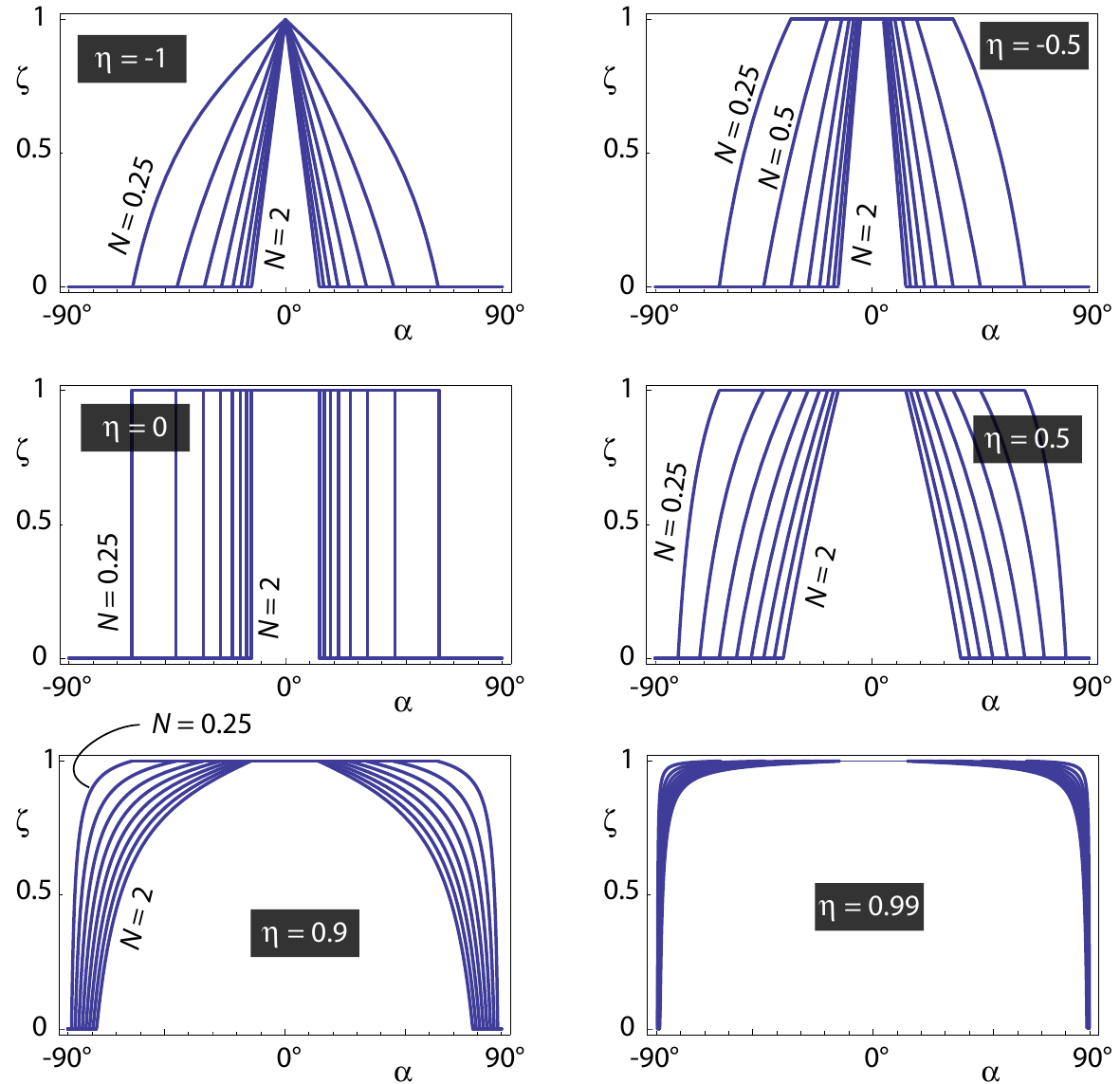} \end{center}
\caption{\label{zeta(alpha)-etas-case1-figure}Dependence of $\zeta(\alpha)$ on $N$, plotted for different values of $\eta$ in the range $-1 \leq \eta \leq 1$.
The curves are plotted for $N$ values between $0.25$ and $2$ in steps of $0.25$.}
\end{figure}

Fig.\ \ref{zeta(alpha)-etas-case1-figure} shows a number of such curves, plotted for different values $N$ and $\eta$.
It is possible to observe compatibility with a number of trends in those graphs, for example the monotonic growth of $\alpha^{c2}$ between $\eta = -1$, where $\alpha^{c2} = 0$, and $\eta = 0$, where $\alpha^{c2} = \arctan (1/2 N)$
while $\alpha^{c3}$ stays constant for any fixed value of $N$;
the equality of $\alpha^{c2}$ and $\alpha^{c3}$ at $\eta = 0$;
and the monotonic growth of $\alpha^{c3}$ between $\eta = 0$ and $\eta = 1$ while $\alpha^{c2}$ remains constant.

Another --- not particularly surprising, but nevertheless important --- trend that can be seen in Fig.\ \ref{zeta(alpha)-etas-case1-figure} is the narrowing of the $\zeta(\alpha)$ curve as $|N|$ increases.
This trend is easily explained:  an increase in the modulus of the f-number corresponds to a decrease in the aperture width of all lenslets, and therefore a decrease in $r$, which makes it easier for light that has passed through one lenslet to miss the corresponding lenslet.
This corresponds to a decrease in the field of view, which we take to be the range of angles between $-\alpha^{c3}$ and $+\alpha^{c3}$, i.e.\ the angle range for which standard refraction occurs.


\subsection{\label{case-2-section}The case $|\eta| \geq 1$} 

Next we consider the case $|\eta| \geq 1$, i.e.\ $|f_1| \leq f_2$, which we treat analogously to the previous case (section \ref{case-1-section}).
Condition (\ref{right-order-condition}) is now satisfied provided that $N \leq 0$ if $\eta \geq 1$, or provided that $N \geq 0$ if $\eta \leq -1$.

\begin{figure}
\begin{center} \includegraphics{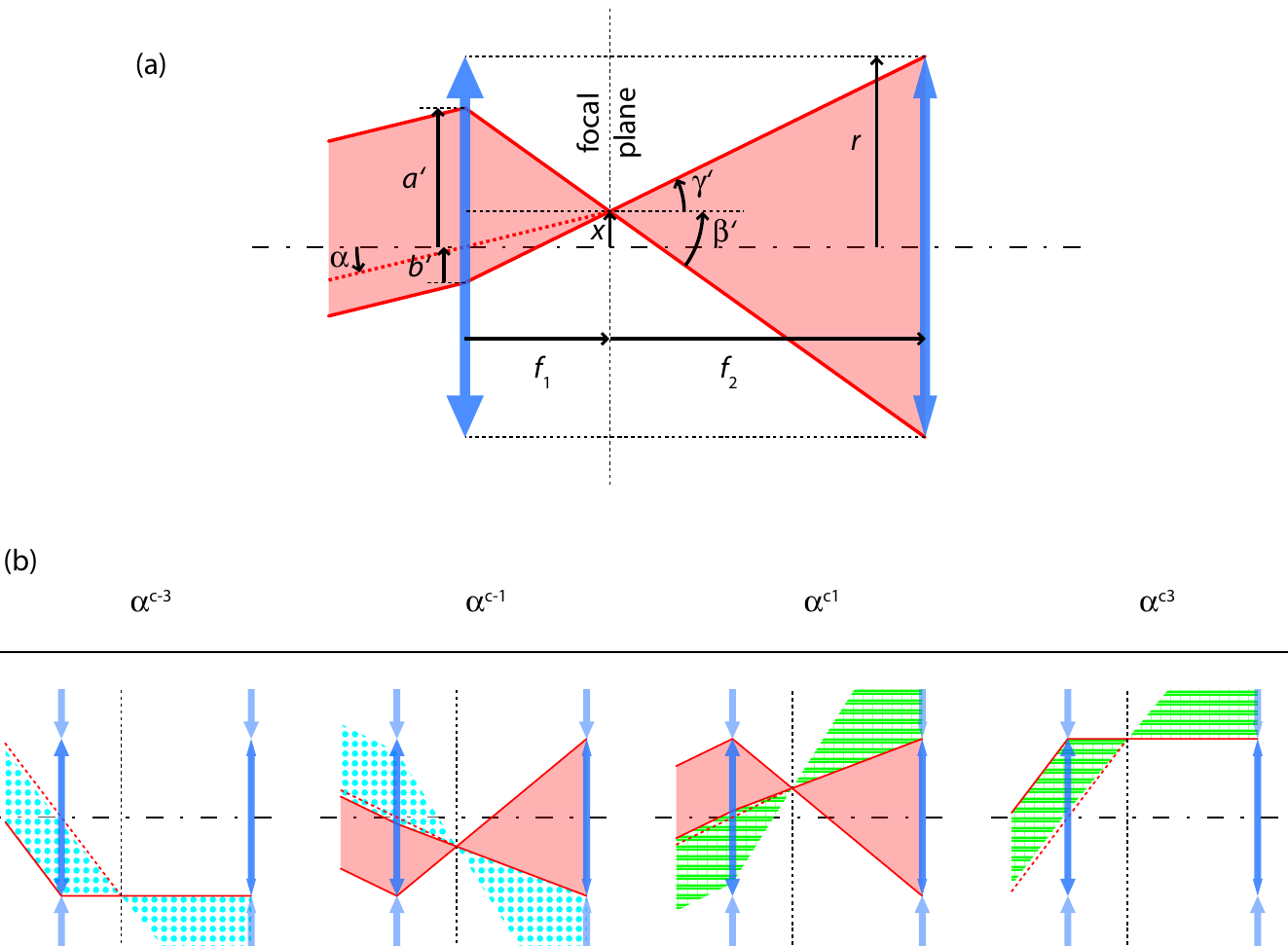} \end{center}
\caption{\label{case-2a-figure}(a)~Geometry of and (b)~critical beam paths through confocal lenslet arrays in the parameter range $\eta \leq -1$.
See the caption of Fig.\ \ref{case-1a-figure} for further details.}
\end{figure}

\begin{figure}
\begin{center} \includegraphics{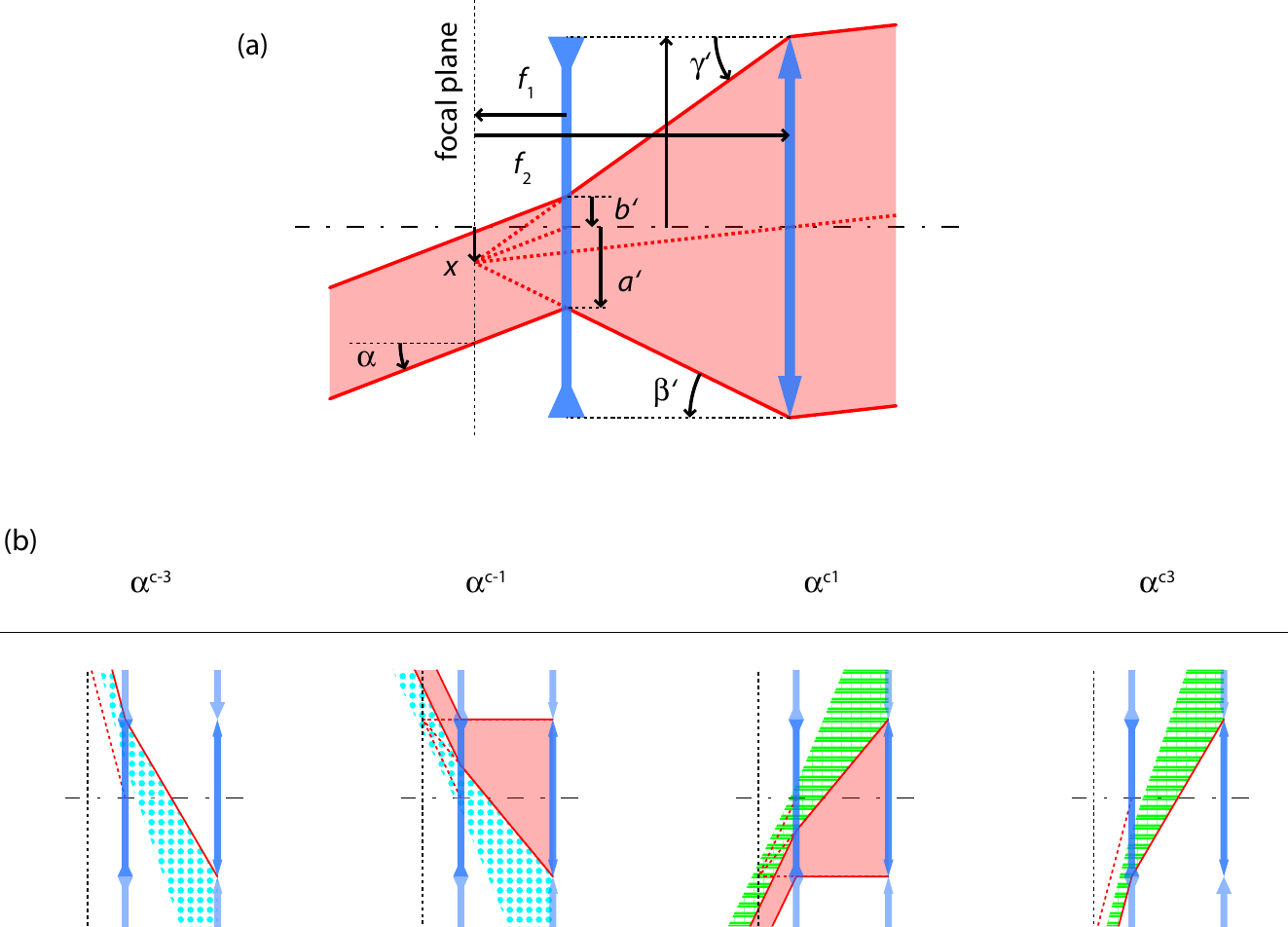} \end{center}
\caption{\label{case-2b-figure}(a)~Geometry of and (b)~critical beam paths through confocal lenslet arrays in the parameter range $\eta \geq 1$.
See the caption of Fig.\ \ref{case-1a-figure} for further details.}
\end{figure}

The key quantities this time are $a^\prime$ and $b^\prime$, which describe the area of the left lenslet through which a light beam with angle of incidence $\alpha$ can enter for all of it to exit through the corresponding right lenslet (Figs \ref{case-2a-figure}(a) and \ref{case-2b-figure}(a)).
We derive expressions for these quantities from the equations
\begin{align}
\tan \gamma^\prime = \frac{b^\prime + x}{f_1} = \frac{r - x}{f_2}, \qquad
\tan \beta^\prime = \frac{a^\prime - x}{f_1} = \frac{r + x}{f_2},
\end{align}
and with $x = f_1 \tan \alpha$ (as before) we find
\begin{align}
a^\prime
&= \frac{f_1}{f_2} [r + (f_1 + f_2) \tan \alpha], 
\label{a-prime-alpha-case-2-eqn} \\
b^\prime
&= \frac{f_1}{f_2} [r - (f_1 + f_2) \tan \alpha].
\label{b-prime-alpha-case-2-eqn}
\end{align}

Starting from normal incidence ($\alpha = 0^\circ)$ and increasing the angle of incidence, what happens qualitatively?
For normal incidence, $a^\prime = b^\prime = r f_1 / f_2$.
In analogy to the previous case, $|a^\prime|, |b^\prime| < r$, but unlike in the previous case this now implies that \emph{not} all the light that enters through the left lenslet exits through the corresponding right lenslet:  only the light rays that enter the left lenslet in the shaded region (which, for $\alpha = 0^\circ$, is centred on the optical axis) exit through the corresponding right lenslet.
As $\alpha$ is increased, the part of the left lenslet through which light rays have to enter to then exit through the corresponding lenslet moves upwards, but does not change in size until one of its edges (the upper side in Fig.\ \ref{case-2a-figure}(b), the lower side in Fig.\ \ref{case-2b-figure}(b)) reaches the edge of the lenslet.
The angle for which it happens is defined as the first critical angle of incidence, $\alpha^{c1}$ \cite{Courtial-2009}.
If $\eta \leq -1$, it can be calculated from the condition $a^\prime = r$ (see Fig.\ \ref{case-2a-figure}(b)), giving
\begin{align}
\tan \alpha^{c1}
= - \frac{r}{f_1} \frac{f_1 - f_2}{f_1 + f_2}
= - \frac{1}{2 N} \frac{1 + \eta}{1 - \eta}
\qquad (\eta \leq -1).
\end{align}
If $\eta \geq 1$, the relevant condition is $a^\prime = -r$ (see Fig.\ \ref{case-2b-figure}(b)), and so
\begin{align}
\tan \alpha^{c1}
= - \frac{r}{f_1}
= - \frac{1}{2 N}
\qquad (\eta \geq 1).
\end{align}

As the angle of incidence is increased further, the part of the lenslet through which light rays have to enter to exit through the corresponding lenslet decreases in size until its size is zero.
If $\eta \leq -1$, this happens when $b^\prime = -r$ (see Fig.\ \ref{case-2a-figure}(b)), i.e.\ when
\begin{align}
\tan \alpha^{c3}
= \frac{r}{f_1}
= \frac{1}{2 N}
\qquad (\eta \leq -1);
\end{align}
if $\eta \geq 1$, the condition is $b^\prime = r$ (see Fig.\ \ref{case-2b-figure}(b)), and so
\begin{align}
\tan \alpha^{c3}
= \frac{r}{f_1} \frac{f_1 - f_2}{f_1 + f_2}
= \frac{1}{2 N} \frac{1 + \eta}{1 - \eta}
\qquad (\eta \geq 1).
\end{align}

As before, negative values of $\alpha$ behave like their positive counterparts with the same modulus.
The relevant critical angles (see Figs \ref{case-2a-figure}(b) and \ref{case-2b-figure}(b)) are
\begin{align}
\alpha^{c-1} = - \alpha^{c1}, \quad \alpha^{c-3} = - \alpha^{c3}.
\end{align}

The power fraction of a uniform plane-wave beam that passes through corresponding lenslets is now the width of the left lenslet through which the beam passes, divided by the width of the lenslet.
For $\alpha^{c-1} \leq \alpha \leq \alpha^{c1}$, this fraction is $|a^\prime + b^\prime| / (2 r) = |f_1 / f_2| = |\eta|^{-1}$; for $\alpha \leq \alpha^{c-3}$ and $\alpha \geq \alpha^{c3}$, it is zero.
In between, i.e.\ for $\alpha^{c-3} \leq \alpha \leq \alpha^{c-1}$ and $\alpha^{c1} \leq \alpha \leq \alpha^{c3}$, it is given by
\begin{align}
\zeta
&= \frac{f_2 + f_1}{2 f_2} - \frac{f_1 (f_1 + f_2)}{2 r f_2} \tan |\alpha| 
= \frac{\eta-1 - N (\eta-1) \tan |\alpha|}{2 \eta}
\qquad (\eta \leq -1), \\
\zeta
&= \frac{f_2 - f_1}{2 f_2} + \frac{f_1 (f_1 + f_2)}{2 r f_2} \tan |\alpha|
= \frac{\eta+1 + N (\eta-1) \tan |\alpha|}{2 \eta}
\qquad (\eta \geq 1).
\end{align}
In summary,
\begin{align}
\zeta
= \left\{
\begin{array}{cl}
|\eta|^{-1} & \mbox{if } |\alpha| \leq \alpha^{c1} \\
\frac{ \eta+\sgn(\eta) \left[ 1 + N (\eta-1) \tan |\alpha| \right] }{2 \eta} & \mbox{if } \alpha^{c1} \leq |\alpha| \leq \alpha^{c3} \\
0 & \mbox{if } \alpha^{c3} \leq |\alpha|.
\end{array}
\right.
\label{zeta-case2-equation}
\end{align}

\begin{figure}
\begin{center} \includegraphics{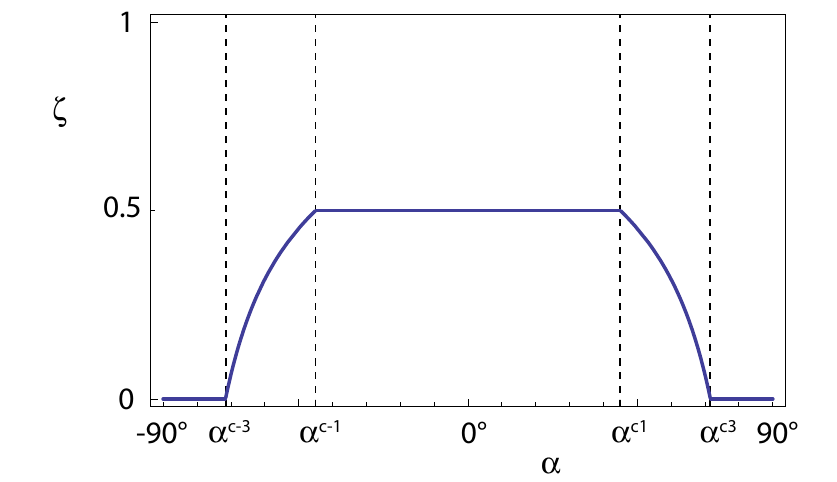} \end{center}
\caption{\label{zeta-alpha-graph-case2-example-figure}Example of a plot of $\zeta$ as a function of $\alpha$ for the case $|\eta| \geq 1$, here $\eta = 2$ and $N = -0.5$.
The critical angles are $\alpha^{c1} = 45^\circ$ and $\alpha^{c3} = 71.6^\circ$.}
\end{figure}

Fig.\ \ref{zeta-alpha-graph-case2-example-figure} shows an example of a plot of $\zeta$ as a function of $\alpha$, calculated according to Eqn (\ref{zeta-case2-equation}). 
The main difference with the corresponding plot for the case $|\eta| \leq 1$ (Fig.\ \ref{zeta-alpha-graph-case1-example-figure}) is that the maximum value $\zeta$ reaches is not 1, but $1/|\eta|$.
As before, the role of the critical angles, this time $\alpha^{c1}$ and $\alpha^{c3}$ is apparent in the graph.

\begin{figure}
\begin{center} \includegraphics{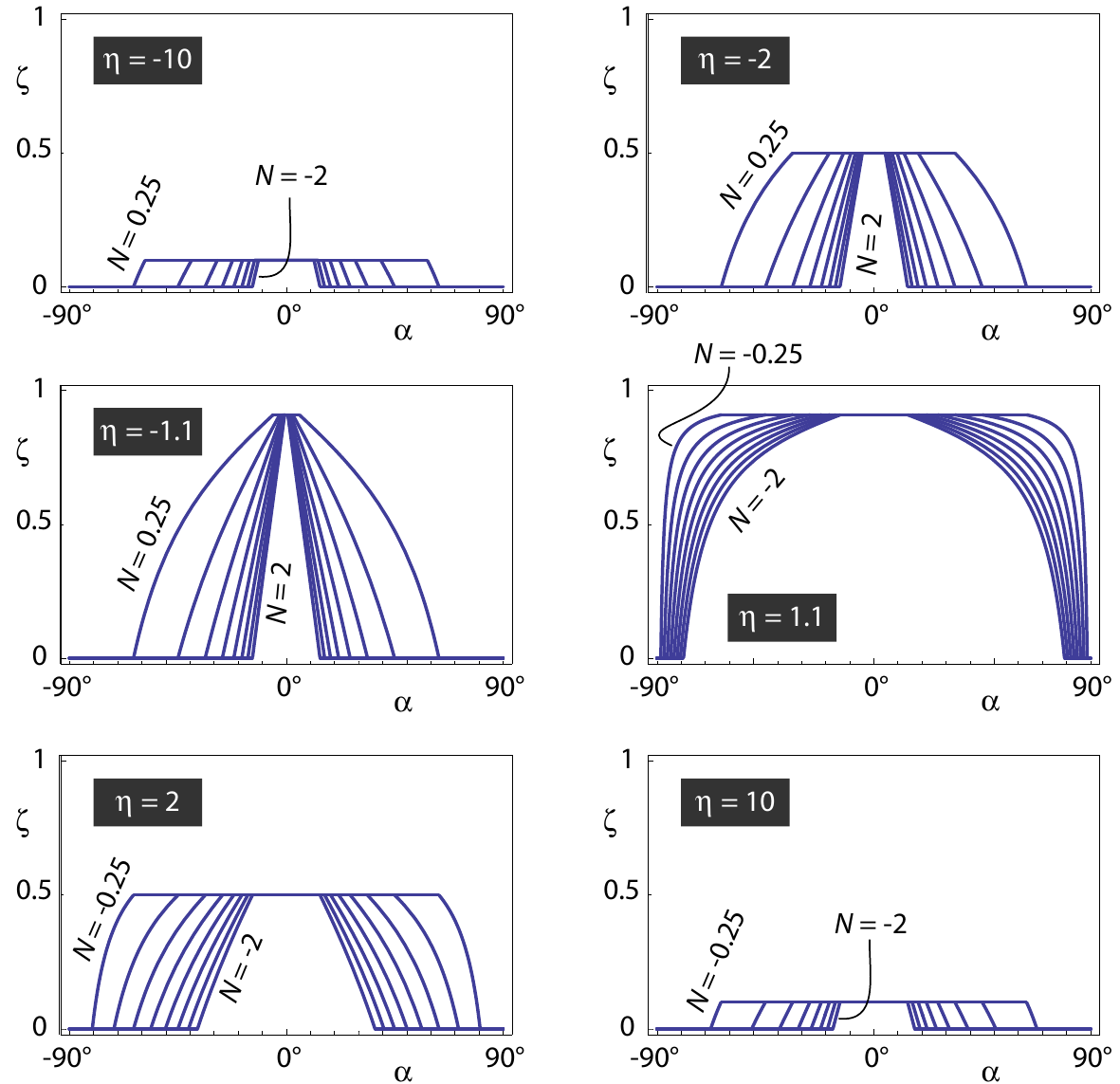} \end{center}
\caption{\label{zeta(alpha)-etas-case2-figure}Dependence of $\zeta(\alpha)$ on $N$, plotted for different values of $\eta$ in the range $|\eta| \geq 1$.
The curves are plotted for $N$ values between $0.25$ and $2$ for $\eta \leq -1$, and between $-0.25$ and $-2$ for $\eta \geq 1$, in steps of $0.25$ in both cases.}
\end{figure}

Fig.\ \ref{zeta(alpha)-etas-case2-figure} shows a number of such curves for different values $N$ and $\eta$, grouped together by their value of $\eta$.
Obviously, the maximum value of these curves is the same for values of $\eta$ with opposite signs, and it approaches the value $1$ as $\eta$ approaches $\pm 1$.
The decrease in the field of view, i.e.\ the range of angles between $-\alpha^{c3}$ and $+\alpha^{c3}$, with increasing $|N|$ is again apparent.

\section{Light undergoing non-standard refraction}

We now turn our attention to non-standard refraction~\cite{Courtial-2009}.
Non-standard refraction is defined as the direction change of light that passes through non-corresponding lenslets, i.e.\ lenslets that do not share an optical axis.

\begin{figure}
\begin{center} \includegraphics{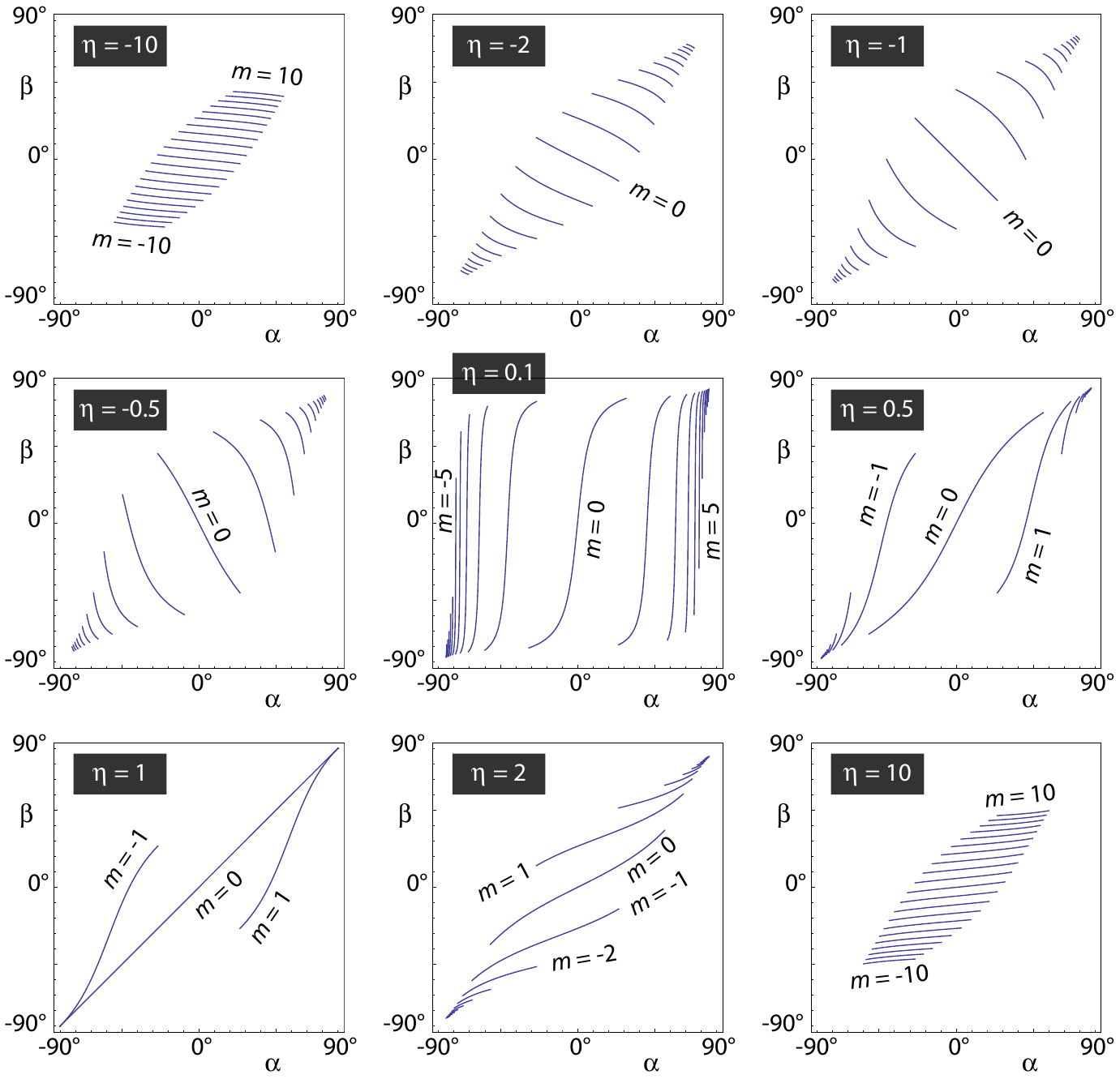} \end{center}
\caption{\label{law-of-refraction-eta-figure}Law of non-standard refraction, Eqn (\ref{law-of-non-standard-refraction-equation}), plotted for various values of $\eta$.
The curves are shown only in the range of incidence angles for which a non-zero fraction of the light exits through non-corresponding lenslets shifted by $m$ lenslets, i.e.\ that between the critical angles $\alpha^{c-3}_m$ and $\alpha^{c3}_m$.
In all cases, $N = 1$; for plots of the dependence on $N$, see Fig.\ \ref{law-of-refraction-N-figure}.}
\end{figure}

\begin{figure}
\begin{center} \includegraphics{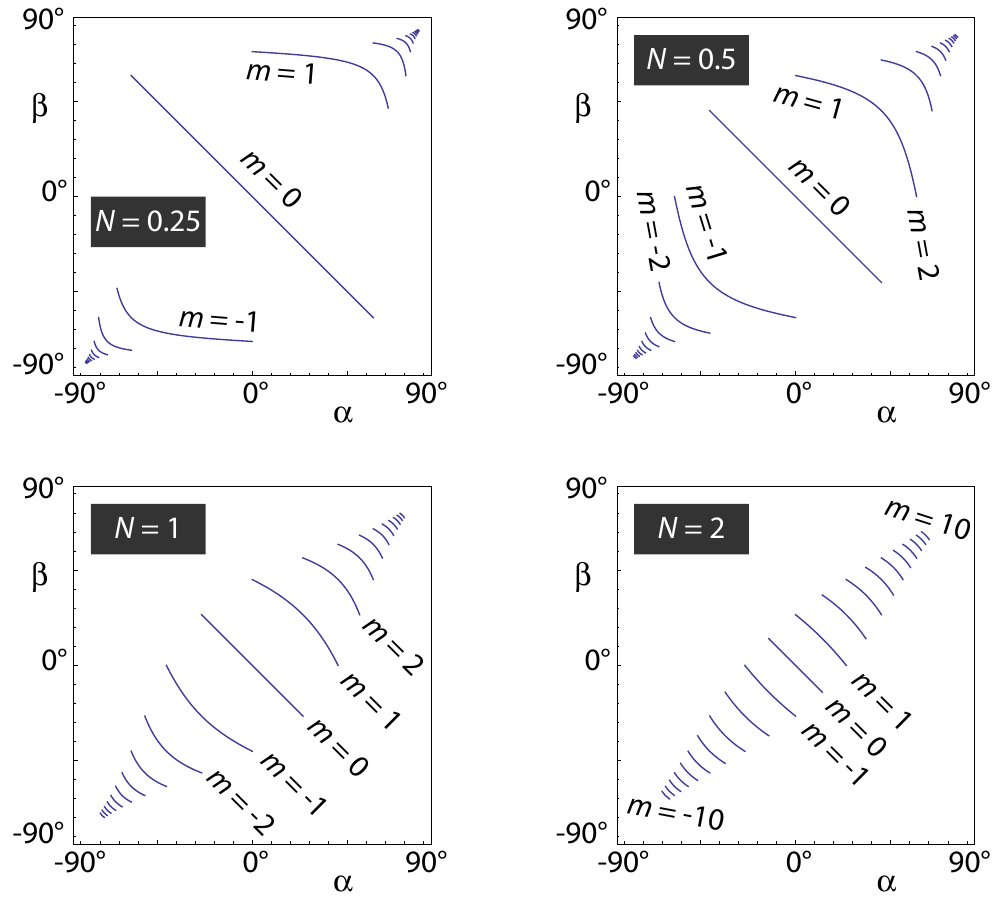} \end{center}
\caption{\label{law-of-refraction-N-figure}Law of non-standard refraction, Eqn (\ref{law-of-non-standard-refraction-equation}), plotted for $\eta = -1$ and different values of $N$.
Like in Fig.\ \ref{law-of-refraction-eta-figure}, the curves are plotted only between the critical angles $\alpha^{c-3}_m$ and $\alpha^{c3}_m$.}
\end{figure}

First we derive the law of refraction for non-standard refraction.
We can do this by treating it a special case of refraction due to generalised CLAs \cite{Hamilton-Courtial-2009b}:  arrays of lenslets in which the telescopes formed by pairs of corresponding lenslets have been modified, but always such that they continue to share a common focal plane.
Suitable modifications can include, for example, a sideways translation of one of the lenslets.
In Ref.\ \cite{Hamilton-Courtial-2009b}, such a translation is defined in terms of a dimensionless parameter $\delta = d / f_1$, where $d$ is the translation distance.
If the sideways translation is the only generalisation of standard CLAs (Fig.\ \ref{CLAs-figure}), then the law of refraction describing transmission through the CLAs is $\tan \alpha = \delta + \eta \tan \beta$.
As expected, for $\delta = 0$, this equation becomes the law of refraction for standard refraction, Eqn (\ref{tan-refraction-equation}).
As a non-corresponding lenslet can simply be seen as the corresponding lenslet, but translated sideways through a distance $d = m 2 r$, where $m$ is an integer, non-standard refraction can be seen as a special case of refraction with generalised CLAs with $\delta = m 2 r / f_1 = m/N$ (and $\theta = 0$, i.e.\ no rotation).
Therefore the law of refraction for non-standard refraction through non-corresponding lenslets is
\begin{equation}
\tan \alpha = \frac{m}{N} + \eta \tan \beta.
\label{law-of-non-standard-refraction-equation}
\end{equation}
Figs \ref{law-of-refraction-eta-figure} and \ref{law-of-refraction-N-figure} show graphs of the angle of refraction, $\beta$, as a function of the angle of incidence, $\alpha$, for a few combinations of $\eta$ and $m/N$.

\begin{figure}
\begin{center} \includegraphics{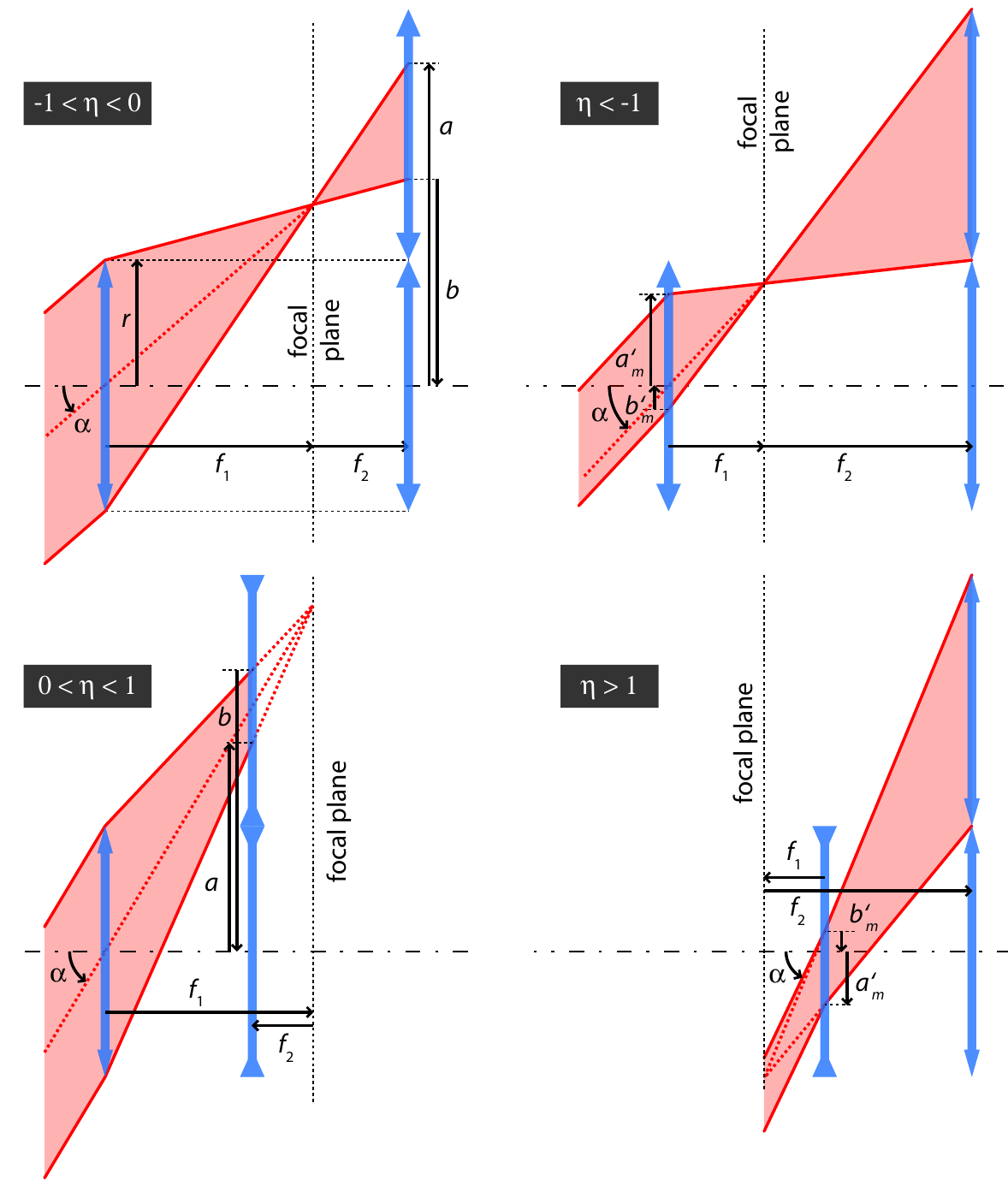} \end{center}
\caption{\label{non-standard-refraction-geometry-figure}Geometry of light undergoing non-standard refraction.
In the examples sketched here, light enters the left lenslet but does not exit through the corresponding lenslet on the same optical axis.
Instead, it exits through the $m$th lenslet above the corresponding lenslet.
The diagrams are drawn for $m = 1$.
}
\end{figure}

Next, we derive the power fraction of incident plane-wave light that undergoes non-standard refraction with a given value of $m$.
We refer to Fig.\ \ref{non-standard-refraction-geometry-figure}, which sketches the geometry of light undergoing non-standard refraction with $n = 1$ in the four cases ($-1\leq \eta \leq 0$, $0 \leq \eta \leq 1$, $\eta \leq -1$, and $\eta \geq 1$) to which Figs \ref{case-1a-figure}, \ref{case-1b-figure}, \ref{case-2a-figure} and \ref{case-2b-figure} refer.
Like in the calculation of the power fraction that undergoes standard refraction (section \ref{standard-refraction-section}), the quantities $a$, $b$, $a^\prime$ and $b^\prime$ (see Fig.\ \ref{non-standard-refraction-geometry-figure}) are key to our current calculation.
$a$ and $b$ are still given by Eqns (\ref{a-alpha-case-1-eqn}), (\ref{b-alpha-case-1-eqn}), but
$a^\prime$ and $b^\prime$ are now dependent on $m$, which we indicate with a subscript $m$:
\begin{align}
a^\prime_m
&= \frac{f_1}{f_2} \left[ (1 - 2 m) r + (f_1 + f_2) \tan \alpha \right], \\
b^\prime_m
&= \frac{f_1}{f_2} \left[ (1 + 2 m) r - (f_1 + f_2) \tan \alpha \right].
\end{align}
For $m = 0$, these expressions reduce to the ones we used in section \ref{case-2-section}, Eqns (\ref{a-prime-alpha-case-2-eqn}) and (\ref{b-prime-alpha-case-2-eqn}), as one would expect.
The calculations in this section generally follow the same outlines as those in sections \ref{case-1-section} and \ref{case-2-section}, so we keep them brief.

The diagrams in Fig.\ \ref{non-standard-refraction-geometry-figure} are drawn for values of the angle of incidence, $\alpha$, that have been chosen such that the maximum fraction of the incident light undergoes non-standard refraction with $m=1$.
For $|\eta| \leq 1$, this power fraction is 1; for $|\eta| \geq 1$, it is $1/|\eta|$, as before.

As $\alpha$ is increased, the transmitted power fraction stays constant until the beam hits the top edge of the right lenslet (for $|\eta| \leq 1$) or the left lenslet (for $|\eta| \geq 1$).
This happens when $a = (2m+1) r$ if $-1 \leq \eta \leq 0$;
when $b = - (2m+1) r$ if $0 \leq \eta \leq 1$; 
when $a^\prime_m = r$ if $\eta \leq -1$;
and when $a^\prime_m = -r$ if $\eta \geq 1$.
We take these conditions to define critical angles;
for $|\eta| \leq 1$,
\begin{equation}
\tan \alpha^{c2,m} =
\left\{
\begin{array}{ll}
\frac{1}{2 N} \frac{2m+1+ \eta}{1 - \eta}
& \mbox{if } -1 \leq \eta \leq 0, \\
\frac{1}{2 N} \frac{2 m + 1 - \eta}{1 - \eta}
& \mbox{if } 0 \leq \eta \leq 1;
\end{array}
\right.
\end{equation}
for $|\eta| \geq 1$,
\begin{equation}
\tan \alpha^{c1,m} =
\left\{
\begin{array}{ll}
\frac{1}{2 N} \frac{2 m - 1 - \eta}{1 - \eta}
& \mbox{if } \eta \leq -1, \\
\frac{1}{2 N} \frac{2 m - 1 + \eta}{1 - \eta}
& \mbox{if } \eta \geq 1.
\end{array}
\right.
\end{equation}

As $\alpha$ is increased further, the power fraction reduces until it reaches zero, when
$b = -(2 m + 1) r$ if $-1 \leq \eta \leq 0$;
$a = (2 m + 1) r$ if $0 \leq \eta \leq 1$;
$b^\prime_m = -r$ if $\eta \leq -1$;
$b^\prime_m = r$ if $\eta \geq 1$.
These again define critical angles:
\begin{equation}
\tan \alpha^{c3,m} =
\left\{
\begin{array}{ll}
\frac{1}{2 N} \frac{2 m + 1 - \eta}{1 - \eta} & \mbox{if } -1 \leq \eta \leq 0, \\
\frac{1}{2 N} \frac{2 m + 1 + \eta}{1 - \eta} & \mbox{if } 0 \leq \eta \leq 1, \\
\frac{1}{2 N} \frac{2 m + 1 - \eta}{1 - \eta} & \mbox{if } \eta \leq -1, \\
\frac{1}{2 N} \frac{2 m + 1 + \eta}{1 - \eta} & \mbox{if } \eta \geq 1.
\end{array}
\right.
\end{equation}

Starting again from incidence angles for which the maximum fraction of the incident light undergoes non-standard refraction (angles such as the ones for which Fig.\ \ref{non-standard-refraction-geometry-figure} is drawn), we now briefly consider what happens when the angle of incidence is decreased.

The power fraction stays constant until the beam hits the bottom edge of the right lenslet (for $|\eta|\leq 1$; see Figs \ref{case-1a-figure}(b) and \ref{case-1b-figure}(b)) or the top edge of the left lenslet (for $|\eta| \geq 1$; see Figs \ref{case-2a-figure}(b) and \ref{case-2b-figure}(b)).
This happens
when $b = -(2 m - 1) r$ if $-1 \leq \eta \leq 0$; 
when $a = (2 m - 1) r$ if $0 \leq \eta \leq 1$; 
when $b^\prime_m = r$ if $\eta \leq -1$; and 
when $b^\prime_m = -r$ if $\eta \geq 1$.
For $|\eta| \leq 1$, we call the angle for which this happens $\alpha^{c-2,m}$, and calculate it to be
\begin{equation}
\alpha^{c-2,m}
\left\{
\begin{array}{ll}
\frac{1}{2 N} \frac{2 m - 1 - \eta}{1 - \eta} & \mbox{if } -1 \leq \eta \leq 0, \\
\frac{1}{2 N} \frac{2 m - 1 + \eta}{1 - \eta} & \mbox{if } 0 \leq \eta \leq 1.
\end{array}
\right.
\end{equation}
As $2 m - 1 = 2 (m-1) + 1$, $\alpha^{c-2,m} = \alpha^{c3,m-1}$ in both cases.
For $|\eta| \geq 1$, we call the relevant angle $\alpha^{c-1,m}$, and we find that
\begin{align}
\alpha^{c-1,m} =
\left\{
\begin{array}{ll}
\frac{1}{2 N} \frac{2 m + 1 + \eta}{1 - \eta} & \mbox{if } \eta \leq -1, \\
\frac{1}{2 N} \frac{2 m + 1 - \eta}{1 - \eta} & \mbox{if } \eta \geq 1.
\end{array}
\right.
\end{align}

As the angle of incidence is decreased further, the power fraction decreases until it reaches zero, when
$a = (2 m - 1) r$ for $-1 \leq \eta \leq 0$; 
$b = -(2 m - 1) r$ for $0 \leq \eta \leq 1$; 
$a^\prime_m = -r$ for $\eta \leq -1$;
$a^\prime_m = r$ for $\eta \geq 1$.
The corresponding critical angle, $\alpha^{c-3,m}$, is then
\begin{equation}
\alpha^{c-3,m} =
\left\{
\begin{array}{cl}
\frac{1}{2 N} \frac{2 m - 1 + \eta}{1 - \eta} & \mbox{if } -1 \leq \eta \leq 0, \\
\frac{1}{2 N} \frac{2 m - 1 - \eta}{1 - \eta} & \mbox{if } 0 \leq \eta \leq 1, \\
\frac{1}{2 N} \frac{2 m - 1 + \eta}{1 - \eta} & \mbox{if } \eta \leq -1, \\
\frac{1}{2 N} \frac{2 m - 1 - \eta}{1 - \eta} & \mbox{if } \eta \geq 1.
\end{array}
\right.
\end{equation}

Finally, we are ready to calculate the power fraction of a uniform plane wave incident at an angle $\alpha$ that enters through a lenslet in the left array and exits through the lenslet in the right array that is $m$ lenslets above that corresponding to the entrance lenslet.
We call this power fraction $\zeta_m$.

We start with the easy cases.
For angles of incidence, $\alpha$, below $\alpha^{c-3}_m$ and above $\alpha^{c3}_m$, $\zeta_m = 0$ irrespective of the value of $\eta$.
If $|\eta| \leq 1$, then $\zeta_m = 1$ for  $\alpha^{c-2}_m \leq \alpha \leq \alpha^{c2}_m$;
if $|\eta| \geq 1$, then $\zeta_m = 1 / |\eta|$ for $\alpha^{c-1}_m \leq \alpha \leq \alpha^{c1}_m$.

The remaining cases are slightly more complicated, and there are a number of them.
They can be calculated from the following expressions.
\begin{enumerate}
\item For $-1 \leq \eta \leq 0$,
$\zeta_m = [(2 m + 1) r + b] / (a+b)$ if $\alpha^{c2}_m \leq \alpha \leq \alpha^{c3}_m$ and
$\zeta_m = [-(2 m - 1) r + a] / (a+b)$ if $\alpha^{c-3}_m \leq \alpha \leq \alpha^{c-2}_m$;
\item for $0 \leq \eta \leq 1$,
$\zeta_m = [(2 m + 1) r - a] / (-a-b)$ if $\alpha^{c2}_m \leq \alpha \leq \alpha^{c3}_m$ and
$\zeta_m = [-(2 m - 1) r - b] / (-a-b)$ if $\alpha^{c-3}_m \leq \alpha \leq \alpha^{c-2}_m$;
\item for $\eta \leq -1$,
$\zeta_m = [r + b^\prime_m] / (2 r)$ if $\alpha^{c2}_m \leq \alpha \leq \alpha^{c3}_m$ and
$\zeta_m = [r + a^\prime_m] / (2 r)$ if $\alpha^{c-3}_m \leq \alpha \leq \alpha^{c-2}_m$;
\item and for $\eta \geq 1$,
$\zeta_m = [r - b^\prime_m] / (2 r)$ if $\alpha^{c2}_m \leq \alpha \leq \alpha^{c3}_m$ and
$\zeta_m = [r - a^\prime_m] / (2 r)$ if $\alpha^{c-3}_m \leq \alpha \leq \alpha^{c-2}_m$.
\end{enumerate}
The results can be summarised as follows.
For $-1 \leq \eta \leq 1$,
\begin{equation}
\zeta_m = \left\{
\begin{array}{cl}
0 & \mbox{ if } \alpha \leq \alpha^{c-3}_m , \\
\frac{ \eta + \sgn (\eta) \left[ - 1 + 2 m - 2( 1-\eta ) N \tan \alpha \right]}{2 \eta} & \mbox{ if } \alpha^{c-3}_m \leq \alpha \leq \alpha^{c-2}_m , \\
1 & \mbox{ if } \alpha^{c-2}_m \leq \alpha \leq \alpha^{c2}_m , \\
\frac{ \eta + \sgn (\eta) \left[ - 1 - 2 m + 2( 1-\eta ) N \tan \alpha \right]}{2 \eta} & \mbox{ if } \alpha^{c2}_m \leq \alpha \leq \alpha^{c3}_m , \\
0 & \mbox{ if } \alpha^{c3}_m \leq \alpha ;
\end{array}
\right.
\label{zeta_m-case1-equation}
\end{equation}
for $|\eta| \geq 1$,
\begin{equation}
\zeta_m = \left\{
\begin{array}{cl}
0 & \mbox{ if } \alpha \leq \alpha^{c-3}_m , \\
\frac{ \eta + \sgn (\eta) \left[ 1 - 2 m + 2( 1-\eta ) N \tan \alpha \right]}{2 \eta} & \mbox{ if } \alpha^{c-3}_m \leq \alpha \leq \alpha^{c-1}_m , \\
\frac{1}{|\eta|} & \mbox{ if } \alpha^{c-1}_m \leq \alpha \leq \alpha^{c1}_m , \\
\frac{ \eta + \sgn (\eta) \left[ 1 + 2 m - 2( 1-\eta ) N \tan \alpha \right]}{2 \eta} & \mbox{ if } \alpha^{c1}_m \leq \alpha \leq \alpha^{c3}_m , \\
0 & \mbox{ if } \alpha^{c3}_m \leq \alpha .
\end{array}
\right.
\label{zeta_m-case2-equation}
\end{equation}

\begin{figure}
\begin{center} \includegraphics{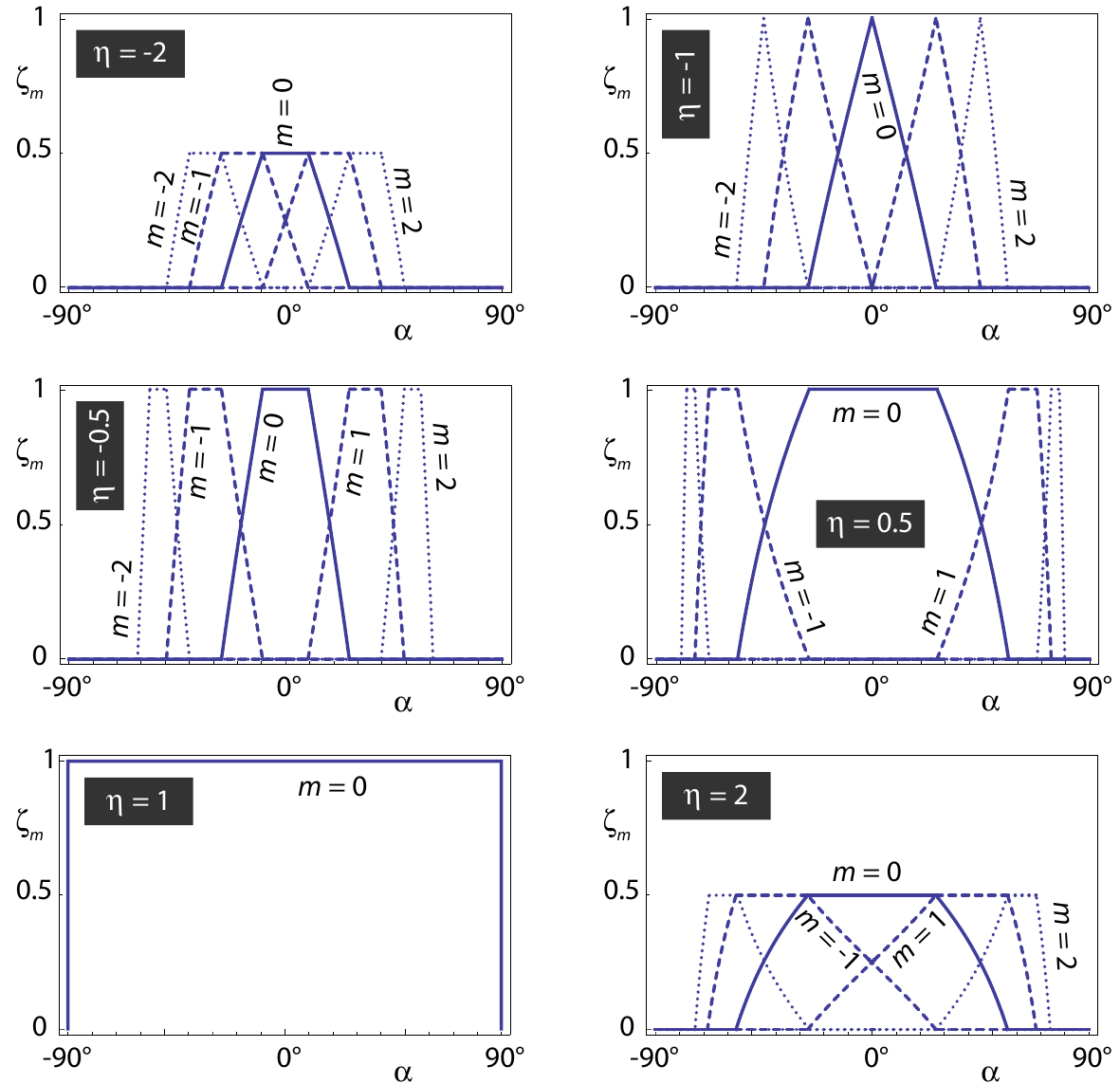} \end{center}
\caption{\label{zeta_m-eta-figure}Power fraction $\zeta_m$ (Eqns (\ref{zeta_m-case1-equation}) and (\ref{zeta_m-case2-equation})), plotted as a function of the angle of incidence, $\alpha$, for $m = 0, \pm1, \pm2$ (respectively represented by lines that are solid, dashed, and dotted) for various values of $\eta$.
All curves were calculated for $|N| = 1$, i.e.\ $N = -1$ for $\eta = 2$ and $N = +1$ for all other cases.}
\end{figure}

\begin{figure}
\begin{center} \includegraphics{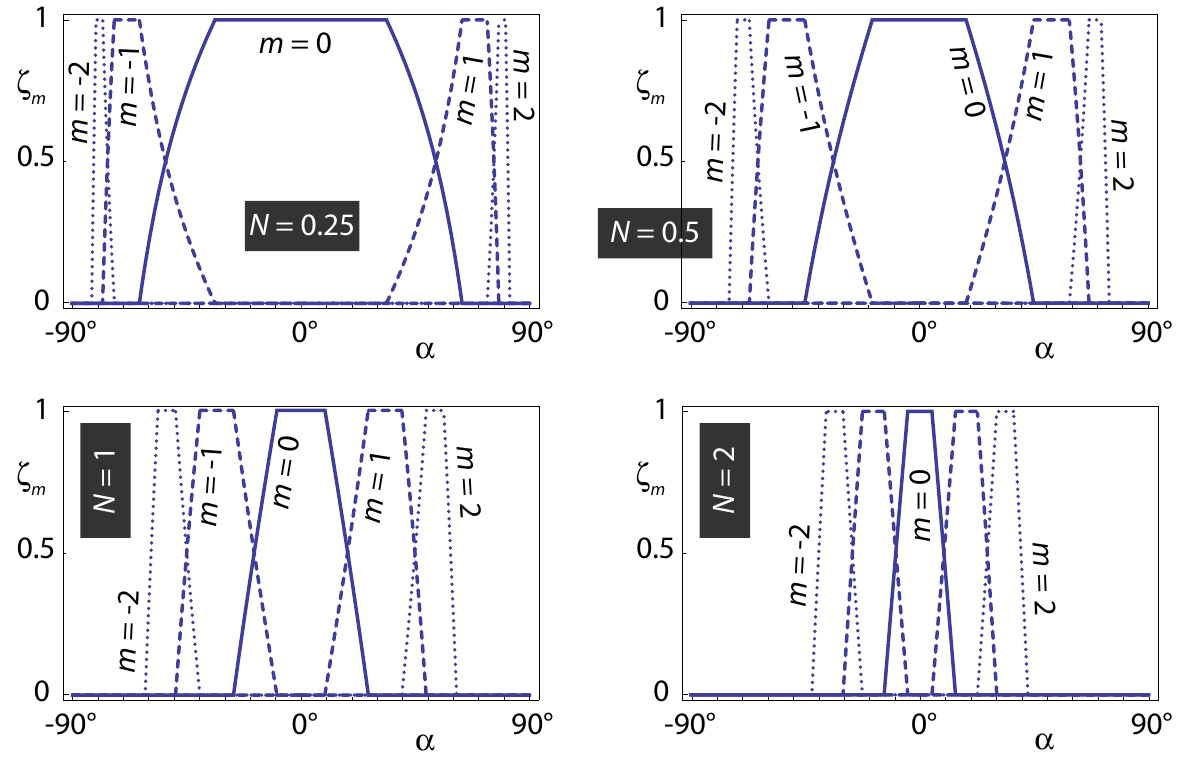} \end{center}
\caption{\label{zeta_m-N-figure}Plots of the power fraction $\zeta_m$ (Eqns (\ref{zeta_m-case1-equation}) and (\ref{zeta_m-case2-equation})) as a function of the angle of incidence, $\alpha$, as $N$ is varied.
Graphs for $m = 0$ (solid lines), $m = \pm1$ (dashed lines) and $m = \pm2$ (dotted lines) are shown.
All graphs were calculated for $\eta = -0.5$.}
\end{figure}

Fig.\ \ref{zeta_m-eta-figure} shows plots of $\zeta_m$ as a function of $\alpha$ for various different values of $\eta$.
A number of things can be seen in those graphs.
Perhaps most notably the curves suggest that the values of $\zeta$ for all values of $m$ add up to one for all values of $\eta$ and all angles of incidence $\alpha$ between $-90^\circ$ and $+90^\circ$, and indeed we confirmed this numerically for a number of cases.
It can also be seen that the peak in the $\zeta_m$ curve moves to greater values of $\alpha$ as $m$ is increased (provided the other parameters are unchanged).

All the curves in Fig.\ \ref{zeta_m-eta-figure} were plotted for $|N| = 1$.
Fig.\ \ref{zeta_m-N-figure} shows how the $\zeta_m$ curves change as $N$ only is varied.
As $|N|$ is increased, the peaks become narrower, and the peaks corresponding to greater values of $|m|$ move inwards.

\section{Similarities with multiple refraction and reflection of cold atoms in light fields}

It is interesting to note that the law of refraction for light passing through non-corresponding lenslets, Eqn (\ref{law-of-non-standard-refraction-equation}), allows a link to recent work on cold atoms.
This works as follows.

With a suitable choice of light fields, the center-of-mass motion of colds atoms can be described by the non-Abelian vector potential term proportional to the spin-$1/2$ operator \cite{Stanescu-et-al-2007,Vaishnav-et-al-2008,Juzeliunas-et-al-2008a,Juzeliunas-et-al-2008b}.
The atomic motion is then characterised by two dispersion branches containing the areas of both positive and negative slopes.
In the low-energy part of the spectrum for each frequency, there are two wave vectors corresponding to the positive and negative group velocity, respectively.
As a consequence, the incident atomic wave may split into two reflected waves at a barrier, one that undergoes specular reflection, and an additional wave that undergoes non-specular reflection \cite{Juzeliunas-et-al-2008b}.
Note that a similar kind of double reflection can occur also for electrons affected by the Rashba spin-orbit coupling \cite{Dargys-2010,Teodorescu-Winkler-2009}.
On the other hand, the transmitted wave experiences the negative refraction for small angles of incidence \cite{Juzeliunas-et-al-2008a}, whereas for large angles of incidence an additional positive refraction takes place \cite{Juzeliunas-et-al-2010}.
The former negative refraction is similar to that taking place by passing the light through the corresponding lenses, whereas the latter ordinary refraction corresponds to the light passing through  non-corresponding lenslets.

\section{Conclusions}

We have calculated the fraction of optical power that undergoes standard refraction on transmission through CLAs.
The calculation requires different cases to be treated separately, making it slightly cumbersome.
The fruit of this drawn-out labour is a number of equations describing the transmission of optical power through CLAs, which can be used for assessing the suitability of CLAs for potential applications.

But the fraction of power transmitted through CLAs is only one of a number of considerations in potential applications.
An example of an additional consideration is the accuracy with which CLAs redirect light rays, which is determined by how close the effect of each individual lenslet is to that of an ideal thin lens.
As lenslet arrays are usually arrays of very basic lenses (typically bumps of glass or plastic), they work best for rays passing close to the centre, i.e.\ for high f-number.
Our results show (not very surprisingly) that a CLA's field of view is lower for higher f-numbers, and so a compromise needs to be found that suits a particular application.
Despite these complications, we believe that our results will be key to investigating the suitability of CLAs for different potential applications, specifically solar concentrators.



\end{document}